# Microstructure and Failure Characteristics of Nanostructured Molybdenum–Copper Composites

*Katharina T. Schwarz, Julian M. Rosalie, Stefan Wurster, Reinhard Pippan, and Anton Hohenwarter\**

Liquid-metal infiltrated $Cu_{30}Mo_{70}$ (wt%) is subjected to severe plastic deformation using high-pressure torsion. The initially equiaxed dual-phase structure is gradually transformed into a lamellar structure composed of individual Cu and Mo layers. The thickness of the lamellae varies between the micro- and nanometer ranges depending on the amount of applied strain. Consistent with the refinement of the microstructural features, strength and hardness substantially increase. In addition, an acceptable ductility is found in the intermediate deformation range. An assessment of the damage tolerance of the produced composites is performed by measuring the fracture toughness in different crack propagation directions. The results indicate the development of a pronounced anisotropy with increasing degree of deformation which is an effect of the concurrent alignment of the nanostructured lamellar composite into the shear plane.

## 1. Introduction

Over the last 30 years, a large variety of significant research areas, ranging from microstructural evolution[1,2] through the characterization of important mechanical and functional properties[3,4] to the invention of novel and industrial relevant processing techniques,[5–7] have developed in the SPD (severe plastic deformation) community. Especially in the last 10 years, the fracture behavior of ultrafine-grained (UFG) and nanocrystalline (NC) metals processed by SPD techniques has come to the center of experimental interest in several research groups.[8–15] It was found that UFG and NC metals show a wide spectrum of fracture properties depending on factors such as the testing orientation, extent of SPD deformation, and temperature. The need to test different orientations in SPD-processed materials, which are essential for materials exhibiting pronounced grain aspect ratios, showcased the ability to obtain one of the highest levels of damage tolerance, achievable in technically relevant metallic alloys.[16] At the same time, it needed to be conceded that a simple grain-size-dependent description of the fracture properties in these types of materials, such as the fracture toughness, is widely misleading due to the evolving anisotropy of the fracture resistance, with very strong and weak crack propagation directions being present for the same material state. The inclusion of the testing temperature into the considerations of the failure process proved that the occurrence of weak specimen orientations cannot be simply referred to excessive amounts of impurity levels at the grain boundaries provoking intergranular failure.[17] Nevertheless, a principal classification of the materials into "ductile" and "brittle" ones can be often made following the behavior that would be expected from the coarse-grained counterparts.[18] To conclude, a lot of insight into failure of SPD metals has been be gained. Most of the investigations, however, were performed on pure or single-phase materials.[18] One interesting question that arises is therefore, how dual-phase materials behave, which become increasingly more a focus of research in the SPD community.

In this study, the fracture behavior of liquid-metal infiltrated $Cu_{30}Mo_{70}$ (wt%) was investigated. Molybdenum–copper composites are used in high-power electrical applications, such as heatsinks,[19,20] electrical contacts and microelectronic packaging,[19–22] and high-power microwave devices.[20] For such applications, high electrical and high thermal conductivity are essential properties.[19–24] Molybdenum is ideally suited for microelectronic packaging because its thermal expansion coefficient is close to the one of Si.[21] In a former study where $Cu_{30}Mo_{70}$ was subjected to high pressure torsion (HPT), it could be already shown that a nanostructured, oriented microstructure with hardness levels up to 600 HV is achievable,[25] which would be also well suited for thermoelectric applications. The idea for a continuation of the studies on this alloy is twofold: on the one hand, UFG copper has shown to have good damage tolerance[26] whereas

K. T. Schwarz,[+] Dr. J. M. Rosalie, Dr. S. Wurster, Prof. R. Pippan
Erich Schmid Institute of Materials Science
Austrian Academy of Sciences
Jahnstraße 12, Leoben 8700, Austria

Dr. A. Hohenwarter
Department of Materials Science, Chair of Materials Physics
Montanuniversität Leoben
Jahnstraße 12, Leoben 8700, Austria
E-mail: anton.hohenwarter@unileoben.ac.at

The ORCID identification number(s) for the author(s) of this article can be found under https://doi.org/10.1002/adem.201900474.

[+]Present address: Hendrickson Commercial Vehicle Systems Europe GmbH, Gussstahlwerkstraße 21, Judenburg A-8750, Austria

Mo is known to become severely brittle in the SPD state.[27] The present alloy type provides the possibility to analyze the composite behavior of intrinsically brittle and ductile UFG materials and possibly show that forming a composite is a feasible way to improve the fracture behavior of "brittle" UFG materials, such as Mo. On the other hand, the implementation of nanostructured $Cu_{30}Mo_{70}$ for example in high-power electrical applications may also require an adequate damage tolerance which has not been analyzed yet.

## 2. Experimental Section

In the present study, a liquid-metal infiltrated Mo–Cu composite provided by Plansee SE, Austria with a nominal composition of 30 wt% (40 at%) Cu and 70 wt% (60 at%) Mo was investigated, which will be hereafter specified as $Cu_{30}Mo_{70}$. Two types of HPT samples were in use: "Small" disks with a diameter of 8 mm and an approximate thickness of 0.8 mm were sufficient to prepare tensile samples and to investigate the microstructural evolution to very high deformation strains. In addition for fracture experiments, "large" disks having a diameter of 30 mm and an initial thickness of ≈10 mm were in use with the minor disadvantage of a lower maximum useable hydrostatic pressure and therefore lower degree of maximum deformation. Both specimen types were cut from the as-received material. The small disks were subjected to deformation in a quasi-constrained HPT rig with a capacity of 400 kN; the large samples were processed with a larger device with a maximum load capacity of 4000 kN. In both cases, the process was performed at room temperature applying a pressure of 7.6 and 5.7 GPa, and a rotational speed of 0.2 and 0.07 revolutions per min for the small and large samples, respectively. These parameters represented standard values for the machines used. In the case of the small disks, the number of rotations varied between 0.5 and 8, whereas in the case of the larger disks, 2 or 4 rotations were applied. The deformation introduced in the samples can be related to the shear strain $\gamma$

$$\gamma = \frac{2\pi r}{t} n \quad (1)$$

Here $r$ is the distance from the center of the disk, $t$ is the thickness, and $n$ is the number of rotations. From the large HPT disks, compact-tension C(T) fracture-toughness specimens were manufactured (**Figure 1**a) by electrical discharge machining in general accordance with ASTM standard E399.[28] The samples were produced with a width $W$ of ≈5.2 mm, thickness $B$ of ≈2.6 mm, and a crack length $a$ of ≈2.6 mm. Three different specimen orientations were investigated and the orientation was encoded with two letters that indicated the crack plane normal and expected crack propagation direction according to the used coordinate system as shown in Figure 1a. The samples were prenotched with a razor blade polishing technique and prefatigued under compression–compression loading with a stress ratio ≈20 and a stress-intensity-factor range between 10 and 25 $MPa m^{1/2}$ to introduce a sharp precrack into the material. For comparative purposes, the fracture behavior of the prematerial was investigated as well. For that single-edge notched tension (SENT) samples were produced with $W = 20$ mm, $B = 10$ mm, $a = 10$ mm, and a specimen length $S$ of 90 mm. The samples were tested under three-point loading conditions with a support span length of 80 mm to evaluate the $J$-Integral as per ASTM E1820-08 using a multi-specimen approach.[29]

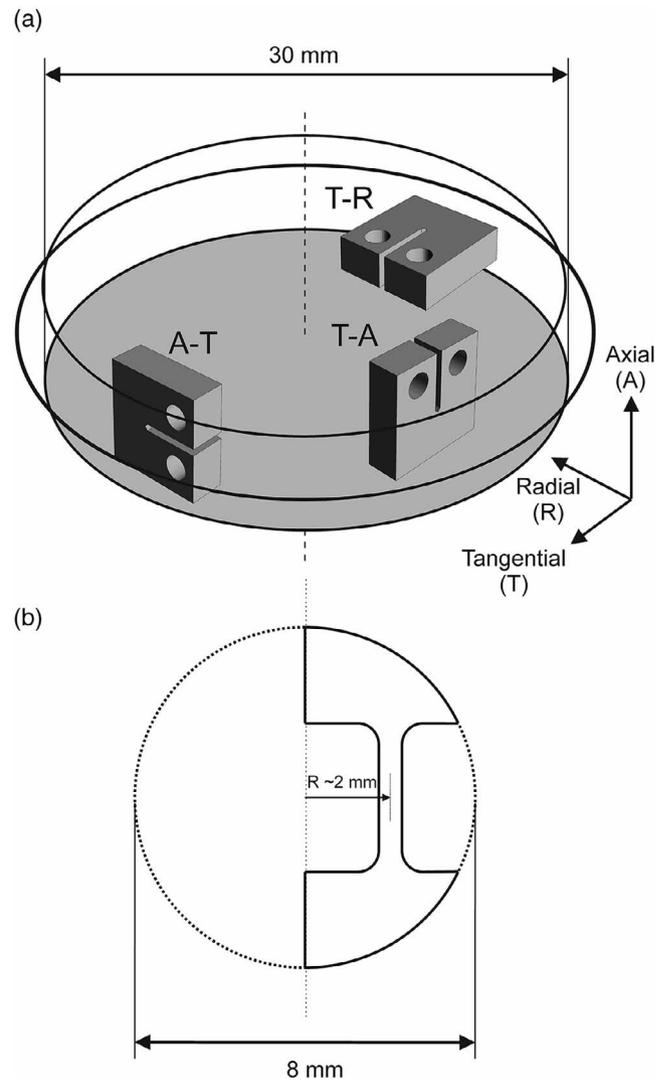

**Figure 1.** a) Schematic of the C(T) samples used for fracture toughness measurements and b) tensile samples.

To assess the tensile properties of the composite, tensile samples with a round cross-section having a diameter of 0.5 mm and a gauge length of 2.5 mm were produced using the small HPT disks. The gauge length was located at a radius of 2 mm (Figure 1b). The tests were performed at room temperature with a tensile testing machine from Kammrath and Weiss equipped with a 2 kN load cell and a crosshead speed of 2.5 $\mu m\,s^{-1}$. For each microstructural condition, one sample was tested. Due to the low number of experiments, an error estimation was carried out with the Gaussian error propagation law. Details to that are summarized in the supporting information to this article. The strain was evaluated using automated digital image correlation. Further details on the specimen production and evaluation process can be found elsewhere.[30]

The hardness of the small samples along the radial direction of the deformed disks was measured with a microhardness tester

Micromet 5104 from Buehler using a load of 9.81 N, a distance of 0.25 mm between individual indents, and a dwell time of 15 s. For the hardness measurements of the large samples, the distance between the indents was taken larger with 1 mm. The hardness of the C(T) samples was measured by making four to six indents close to the notch on both sides of the samples followed by calculating the mean value. These hardness values linked the microstructure and hardness investigated on the small samples with the one to be expected in the large HPT disks. The hardness measurements performed on the large disks can be found in the supporting information to this article. Microstructural and fractographic investigations were performed with a field emission scanning electron microscope (SEM) Zeiss Gemini 1525 possessing an electron back scatter diffraction (EBSD) detector (e⁻Flash$^{FS}$) and an EDS detector (XFlash©6) from Bruker.

## 3. Results

### 3.1. Microstructure Evolution

The undeformed prematerial examined in the tangential direction is shown in **Figure 2** with an inverse pole figure (IPF) (Figure 2a) and a phase map (Figure 2b). It is noted that the same coordinate system for the undeformed as well as for the deformed disks is used (see Figure 1a). The comparison shows that the Mo areas consist of individual grains separated by high-angle boundaries, while in Cu this kind of boundary is rare. This implies that the grain size of the Cu-phase seems to be very large. The average grain size of Mo is ≈6 μm. An inspection into a second viewing direction, for example, the axial direction (not displayed), leads to the same results as regards grain size and the appearance of grain shape. This indicates that the Mo-phase has a spheroidal grain shape however no specific texture. In contrast, the Cu-phase seems to be textured or consist of relatively large grains as a result of the infiltration process.

In **Figure 3**, the most significant steps of structural refinement observed parallel to the tangential direction are presented using a back-scatter electron detector in the SEM. Figure 3a shows the structure after a shear strain, γ, of ≈44. The Mo-phase becomes compressed and elongated in the shear plane. With increasing deformation (γ ∼ 94), Figure 3b, the Mo-phase is continuously refined through a combination of structural elongation and necking. In addition, substructure formation can be observed within the individual phases, as shown in Figure 3c. This substructure formation, which can be compared with the microstructural refinement of the individual pure metals, is supposed to be the main cause for ane increase in hardness. For higher strains, the hardness progressively increases as the structure is further refined, Figure 3d (γ ∼ 143). For the highest achieved degree of deformation (Figure 3e), which is γ ∼ 244, the substructure is nanocrystalline. The grain size of the Mo-phase is not determinable by SEM; however, from Figure 3f, it can be inferred that it is smaller than the one to be expected for the equilibrium or saturation grain size of the pure elements, which is rather in the range of a couple hundreds of nanometers.[31] A more comprehensive investigation of the microstructural evolution of the present alloy can be found elsewhere.[25]

Along with the microstructural refinement, the hardness and tensile behavior was investigated (**Figure 4**). The hardness as a function of applied strain (Figure 4a) was obtained by converting the radial positions of the hardness measurements into γ (Equation (1)). The hardness of the undeformed material is about 125 HV. The lowest hardness levels, taken from the centers of the disks, are however always higher than the value of the undeformed material due to the hardening contribution of the compression deformation introduced before the HPT disk is rotated. Only for the highest number of rotations, the onset of a saturation regime, indicated by a pronounced change in the slope of the hardness curve, seems to appear. This is consistent with other investigated composites, such as pearlitic steels,[32] in which at a high number of rotations, a saturation regime seems to occur. However, in both cases, the maximum hardness is rather restricted by the hardness of the anvils than by the physical reasons. With the change in hardness, a distinctive increase in yield strength and ultimate tensile strength of the composite was also measured as shown in Figure 4b and summarized in **Table 1**.

In comparison with the undeformed material, see Table 1, the ductility is drastically reduced by even 0.5 rotations of HPT; however, it remains more or less stable up to 2.6 rotations. In this context, it has to be emphasized that the data on the undeformed material, which was provided by the manufacturer Plansee SE, originate from samples with a different geometry (length ≈25 mm, width ≈10 mm). As a result, a straightforward comparison between the samples bears some difficulties. Due to the relatively low hardness, attempts to produce miniaturized samples of the as-received material were not successful. Efforts to measure the tensile behavior of the more heavily SPD-deformed samples resulted in a failure of the sample during the preparation process or during the tensile test in the elastic regime and are therefore not presented. This behavior seems to be related to the changes of fracture toughness.

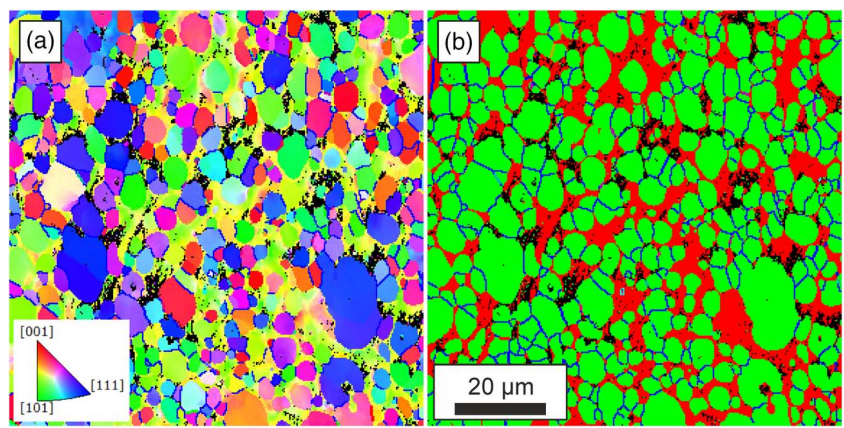

**Figure 2.** a) Representation of the prematerial investigated in the tangential direction with an IPF-map and b) the corresponding phase map. Large angle boundaries are inserted in blue, green areas correspond to Mo, whereas red ones belong to Cu. Black areas represent areas where patterns were unsatisfactorily indexed to display the orientation and phase.

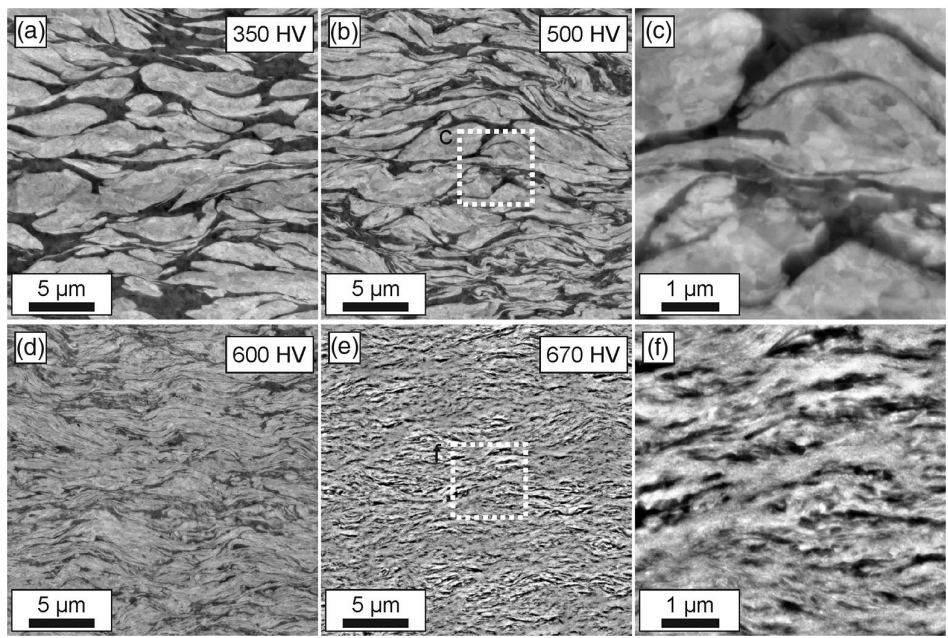

**Figure 3.** Significant changes in the microstructure upon deformation for shear strains of a) $\gamma \sim 44$, b) $\gamma \sim 94$, d) $\gamma \sim 143$, and e) $\gamma \sim 244$ investigated in the tangential direction using large HPT-disks. Magnified views of (b,e) are displayed in (c,f). Mo with the higher atomic number appears bright, while Cu-regions are dark.

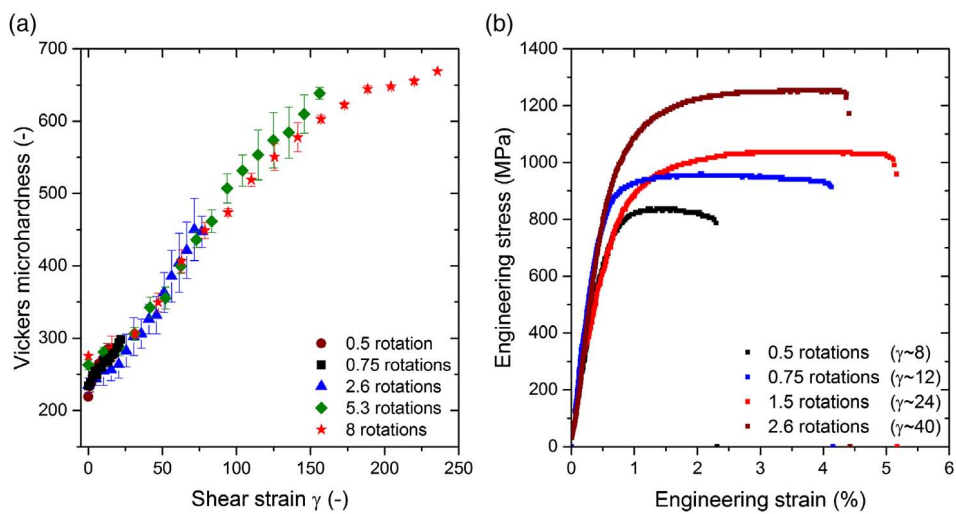

**Figure 4.** Mechanical properties of the investigated composite. a) Hardness evolution as a function of applied shear strain. b) Representative stress–strain curves for different amounts of predeformation. The shear strains are calculated for a radius of 2 mm, where the nominal position of the gauge length is situated.

An examination of the tensile samples' fracture surfaces (**Figure 5**) indicate that in the investigated deformation regime the failure type does not change drastically. For both low degrees of deformation, Figure 5a,b, as well as for higher strains, Figure 5c,d, a microductile fracture surface is recognizable. As the lower magnified images (Figure 5a,c) suggest the extent of necking in the samples seems to be low, which is consistent with the trends shown in the tensile curves, Figure 4b. The remaining ductility of the deformed samples is mainly a consequence of the uniform elongation. Extensive necking, which would be observable as a load decrease in the engineering stress, is not pronounced.

**Table 1.** Yield strength, $\sigma_y$, the ultimate tensile strength, $\sigma_{UTS}$, and ductility in terms of the elongation at fracture, $\varepsilon_{fr}$ of $Cu_{30}Mo_{70}$ subjected to different number of rotations. An error estimation for $\sigma_{UTS}$ yielded a value of ≈2% and 0.1% for the measured strain. For more details, see the supporting information to this article.

| Rotations | $\sigma_y$ [MPa] | $\sigma_{UTS}$ [MPa] | $\varepsilon_{fr}$ [%] |
|---|---|---|---|
| Undeformed | 300 | 482 | 30.5 |
| 0.5 | 790 | 836 | 1.7 |
| 0.75 | 835 | 960 | 3.5 |
| 1.5 | 850 | 1035 | 4.3 |
| 2.6 | 1005 | 1253 | 3.6 |

### 3.2. Fracture Toughness

The fracture toughness of the undeformed state was obtained by measuring the J-Integral according to ASTM-standard E1820.[29] More details to these measurements can be found in the supporting information for this article. A conversion of a J-value to equivalent stress intensity is made by

$$K_{IC} = \sqrt{\frac{J_i E}{(1-\nu^2)}} \tag{2}$$

Here, $K_{IC}$, is the fracture toughness, $E$, the Young's modulus with 226 GPa, $\nu$, is the Poisson ratio, which was assumed to be 0.3, and $J_i$ is the J initiation value, which should not be confused with a $J_{IC}$ value as defined in the ASTM-standard. More details can be found in the supporting information to this article. $J_i$ was measured to be $16\,\mathrm{kJ\,m^{-2}}$, which is equivalent to $K_{IC} = 63\,\mathrm{MPam^{1/2}}$.

A summary of the fracture toughness values presenting the fracture toughness evolution of all three testing directions of the deformed samples as a function of hardness is shown in **Figure 6**. The hardness values are mean values of the hardness (four to six indents) measured on both sides of the specimen near the notch before the test. In addition, the K-value of the undeformed material is inserted. With SPD deformation, which leads to a hardness increase, the fracture toughness generally deteriorates and at the same time a distinctive influence of the testing direction on the fracture resistance becomes recognizable. The A–T orientation contains the lowest values and the T–R orientation contains the highest ones. A comparison of the hardness data in Figure 6 (obtained from the large HPT-disks) with the hardness plots in Figure 4 (measured on the small HPT-disks)

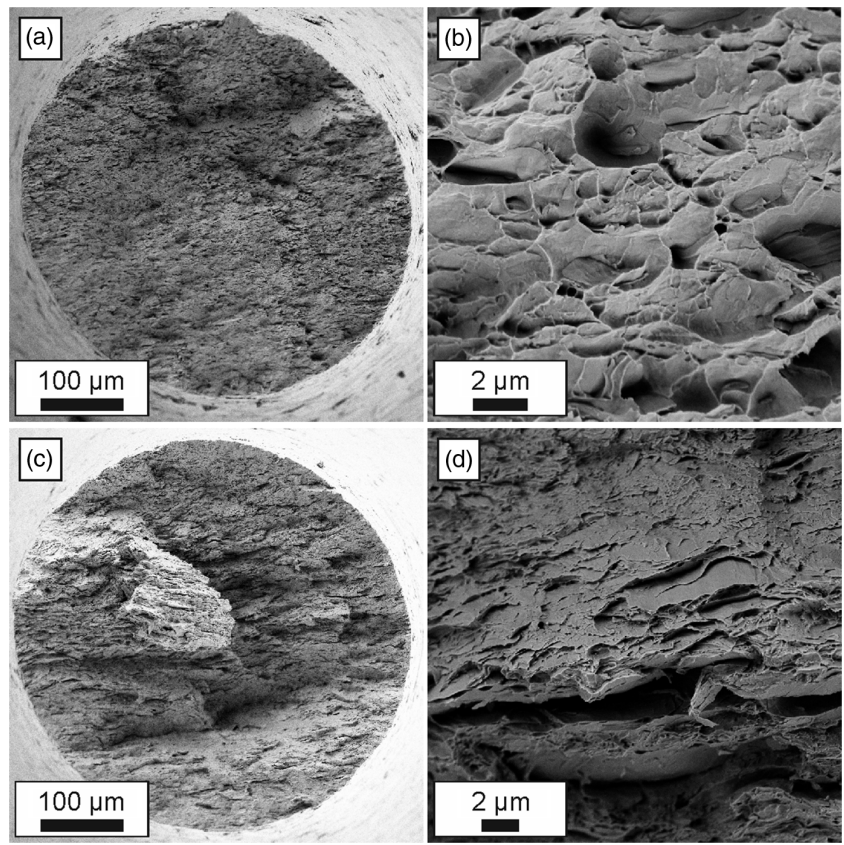

**Figure 5.** Typical fractographs of tensile samples subjected to 0.5 (a,b) and 2.6 (c,d) rotations.

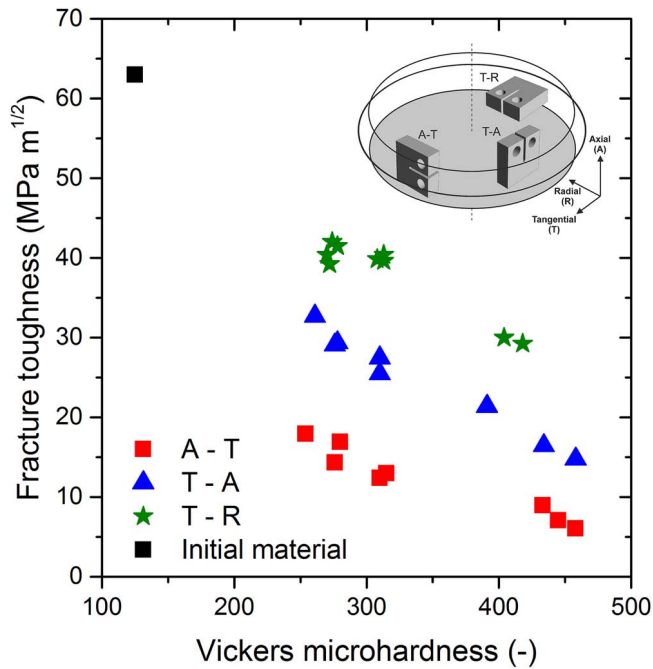

**Figure 6.** Fracture toughness of the $Cu_{30}Mo_{70}$ composite for the different specimen orientations as a function of the hardness of the samples.

## 4. Discussion

The results indicate that the HPT processing of $Cu_{30}Mo_{70}$ leads to substantial changes of the microstructure and mechanical properties. First, the hardness and strength exhibit a strong increase. Second, even though the ductility of the undeformed material is higher, a distinctive amount of ductility can also be measured in the deformed states. Third, the microstructural alignment also leads to a pronounced anisotropy of the fracture resistance which will be in the center of the following discussion.

### 4.1. Anisotropy of Fracture Toughness

To scrutinize the origin for the increasing anisotropy, as shown in Figure 6, typical fractographs of the samples with the different crack propagation directions are shown in **Figure 7**. In the undeformed material at low magnifications, Figure 7a, a relatively rough fracture surface is observable with a distinct transition from the fatigue precrack to the overload fracture surface. At higher magnifications (Figure 7b), the typical mixed fracture type of this composite can be noticed, where the deformation is mainly carried by the ductile Cu-phase whereas in the polyhedral-shaped Mo grains almost no sign of plasticity is observable. In the deformed state, a drastic change in the fractographs is observable depending mainly on the chosen crack growth direction.

In the A–T orientation, the fracture surface becomes gradually flatter with increasing prestrain, Figure 7c, which is a result of the microstructural alignment. This means that the crack path tortuosity is reduced along with the fracture resistance. In addition, the plasticity in the Cu-phase seems to become reduced as shows that the fracture toughness experiments were only performed up to a hardness of ≈450 HV. Problems with specimen preparation due to the low fracture toughness in the A–T orientation are the main cause why specimens with even higher hardness levels were not investigated.

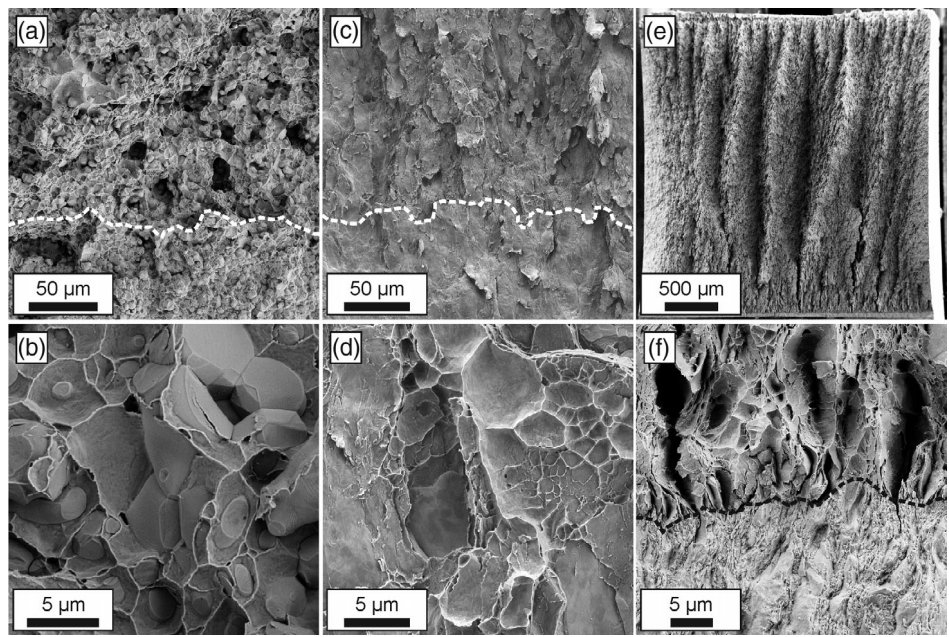

**Figure 7.** Fractographs representing significant fracture features. a,b) Undeformed material (hardness ≈125 HV, fracture toughness ≈63 MPam$^{1/2}$) c,d) deformed material in the A–T orientation (hardness ≈254 HV, fracture toughness ≈18 MPam$^{1/2}$) e,f) T–R orientation (hardness ≈274 HV, fracture toughness ≈42 MPam$^{1/2}$). Dashed lines are inserted where applicable to delineate the transition from the fatigue precrack to the overload fracture surface.

well, which can be inferred from the relatively flat dimple structure, Figure 7d. Due to the re-orientation of the composite and the fact that plasticity during the fracture process is mainly carried by the Cu-phase, void formation leading to the dimple structure is restricted to the aligned copper channels between the Mo-grains. The typical microstructure for the presented fractographs is shown in Figure 3a.

The T–A orientation exhibits the same fracture features as observable in the A–T orientation and are therefore not shown. The reason for the similarities is that in this testing orientation the crack deflects instantly under an angle of ≈90°. This means that the crack propagates in the same crack plane orientation as in A–T samples. The reason for this behavior seems to be a much higher fracture resistance of the material for the precrack being perpendicularly oriented to the aligned microstructure. The fracture toughness values in this direction yield higher values because of the locally deflected crack. In the extreme case of a 90° deflection the crack driving force for the locally kinked crack is only half of the one of a straight propagating crack.[33] Therefore, the apparent fracture toughness in this direction is typically twice the one measured in the A–T orientation where crack deflection does not occur. Nevertheless, these measured values represent only lower bound values. In reality, if it was possible to force the crack to propagate in the desired direction, for example by using deeply side grooved samples, a much higher fracture toughness is expected.

In the T–R orientation, Figure 7e,f, the fracture surface exhibits various delaminations. The positive effect of this feature on keeping the fracture toughness high is well known.[34,35] Their formation prior or at the onset of fracture leads to a change of the stress state at the crack tip from a plane strain- to a plane stress-dominated one. The latter allows more plastic deformation before failure and therefore allows a higher measurable fracture toughness—in this instance, the highest one in a direct comparison of the three investigated specimen orientations. The formation of the delaminations is connected to the results of the A–T orientation. This orientation has the same crack plane and crack opening direction as the delaminations in T–R specimens. Therefore, the local fracture behavior of the secondary cracks, i.e. delaminations, is controlled by the fracture resistance of the A–T specimens and the evolving weak crack path with increasing deformation. This demonstrates that the high fracture resistance in the T–R orientation is a result of having this weak crack growth resistance found in the A–T orientation.

### 4.2. Origin of Weak Crack Path

The origin of the anisotropy seems to be connected with a relatively weak crack path parallel to the tangential direction in which the structural elements are aligned. Analogies to this result can be found in preceding studies dealing with two phase materials such as pearlitic[36] and duplex steels.[37] In case of a pearlitic steel, the interface between ferrite and the partially transformed cementite has been identified as the weakest link. For the duplex steel, intralamellar failure within the ferritic phase as well as interlamellar failure along the ferrite–austenite interface was dominant.[37] To learn more about the prevalent crack path in the A–T orientation with its low fracture toughness, further fractographs of identical areas identified on both fracture specimens halves of a sample with a fracture toughness of 6 MPam$^{1/2}$ (≈460 HV) are shown in **Figure 8**.

In principal, two different morphologies can be observed. First, there are relatively rough areas, which belong to microductile dimple fracture. Second, there are very smooth regions as well, which belongs to a more brittle phase. In addition, the fractographs match each other perfectly or in other words there are no regions observable where on one half a rough feature and on the other half a smooth feature coincide in the same position. EDS measurements (area scans are presented in the supporting information for this article to strengthen the following line of arguments) indicate that the smooth brittle areas consist of Mo and the ductile of Cu (EDS measurements were only taken from one fracture half). This would imply that matching areas belong to the same phase and that intralamellar crack growth, within a phase and not along phase boundaries, prevails. For interlamellar crack growth, one would expect that for example a brittle region on one fracture surface would match with a microductile region on the other fracture surface or vice versa. Only under these circumstances, pure interlamellar failure between a Mo–Cu interface would be expected. However, EDS spot measurements (see **Table 2** in conjunction with Figure 8) and EDS maps in the supporting information also prove that the individual areas consist not only of the respective element. The spot measurements, in which the radius of interaction volume for Mo is in the order of 100 nm (6 kV acceleration voltage), reveal the presence of Cu in the smooth areas and the presence of Mo in the ductile areas. In case of the smooth Mo areas, the Cu is located on top of Mo because in this hardness regime the typical Mo-lamella thickness is in the order of micrometers, see also Figure 3b. Dissolution of Cu in Mo is not expected for the amount of deformation applied, and the microstructural scale either. For the ductile phase, the presence of Mo can be explained by the fact that the Cu-phase during deformation forms very fine lamella resulting in the detection of Mo from an adjacent area beneath. To conclude, the presence of Cu in the Mo-rich areas suggests that intralamellar fracture within the Mo-phase, i.e., intercrystalline fracture, does not occur. The preferential crack path is mainly intralamellar in the Cu-phase and two scenarios are conceivable. i) The crack propagates in the very fine Cu-channels between the Mo-particles leading to the appearance of brittle areas. Without any in-situ experiments, it can however not be judged if the crack runs in the center of the channel or presumably very close to one of the neighboring interfaces. ii) The crack runs in the thicker Cu-lamellae leading to the ductile appearance on the fractograph where resulting dimples are visible on both halves of the fracture surface. These ideas are consistent with the typical microstructure for this fracture toughness regime as shown in Figure 3b,c.

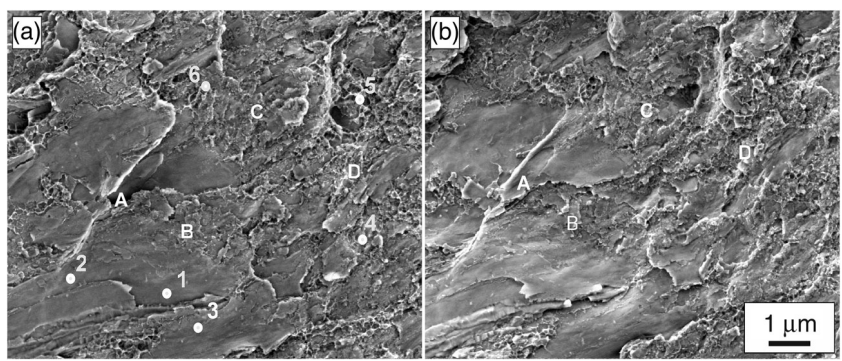

**Figure 8.** Comparison of fracture surfaces taken from identical areas on both halves of the fractured sample with a fracture toughness of 6 MPam$^{1/2}$. It is noted that the left image (a) has been mirrored to facilitate the comparison with the other fracture half (b). Capital letters indicate matching areas on both fracture halves. The numbers indicate locations where EDS-spot measurements were performed.

**Table 2.** EDS spot measurements indicated in Figure 8 executed on both, smooth-brittle and rough-ductile areas. All numbers are in at.%. The precision of the analysis is estimated to be ±1%.

| Position | Mo | Cu |
|---|---|---|
| 1 | 85 | 15 |
| 2 | 86 | 14 |
| 3 | 80 | 20 |
| 4 | 34 | 66 |
| 5 | 11 | 89 |
| 6 | 35 | 65 |

### 4.3. Comparison with Single-Phase Fracture Toughness

The general idea of producing composites is to obtain improved mechanical or functional properties which the single constituents do not possess on their own. To discuss how the fracture toughness in the highly deformed state benefits from this basic principle, it is helpful to take the fracture toughness of the pure elements in the SPD state into account. In the following, only the weak crack path is considered. To measure the fracture toughness of pure HPT-deformed Mo, additional experiments were performed yielding a value of 3.3 MPam$^{1/2}$. The most significant experimental details to this evaluation are summarized in the supporting information to this article. The fracture toughness of pure copper can be found in another publication[26] with a value of 33.4 MPam$^{1/2}$. The large difference in the fracture toughness between the pure elements in the SPD state (≈factor of 10) is also reflected by the typical fractographs, **Figure 9**. SPD-Cu fails in a micro-ductile manner with a pronounced stretched zone, which implies that the crack is strongly blunted before the crack propagates. In contrast, for SPD-deformed Mo, there are almost no indications of plasticity and intergranular failure prevails. At the same time, there is also a large variation in hardness which is ≈140 HV for Cu and 540 HV for Mo.

A simplistic approach to model the fracture toughness of the composite would be to apply a linear rule of mixture. In such a case by considering the volume percentages of Cu (≈33%) and Mo (67%) and the fracture toughness of the pure SPD-processed elements, a fracture toughness of around 13 MPam$^{1/2}$ would be expected, which is twice as high as the lowest one that was measured for the composite. The nonapplicability of this simple rule suggests that for the prediction of the fracture resistance in SPD composites, various other factors need to be taken into account. The main factor is that the intrinsic fracture behavior of pure Cu and Mo differs markedly from the fracture behavior of the composite as shown in Figure 8 and 9. This is a result of the alignment of the structure leading to preferential crack paths, which seem to be intralamellar ones in the Cu-phase. At the same time, intergranular fracture in Mo, as found for the pure metal, does not occur either. The fracture toughness of pure SPD-Cu, which is very high, is however also not controlling the toughness of the composite as the grain size or lamella thickness in the pure

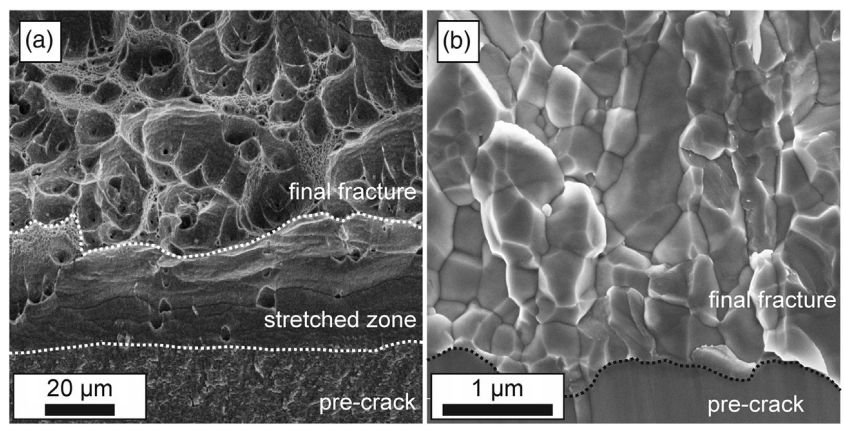

**Figure 9.** Comparison of the failure type of the pure deformed metals. a) SPD-Cu b) SPD-Mo.

SPD-state and within the composite strongly differs. In addition, the deformation of Cu is also strongly constrained between the Mo-phase as a result the same large crack tip opening, as in the case of the pure metal (Figure 9a), is not possible. Other factors which make the prediction of the fracture resistance of the composite even more difficult are that the extent of alignment and as a consequence the fracture toughness also depends on the amount of prestrain. Finally, due to the weak crack path, a pronounced anisotropy in the fracture resistance arises as well. This shows the complexity and problems one has to face in the prediction of fracture properties of such composite structures. Therefore, future work will concentrate on varying the deformation temperature which would have an influence on the microstructural evolution and the resulting fracture toughness anisotropy. At the same time, a post-SPD heat treatment in deformed samples may also have a positive effect on restoring ductility, especially in the Cu-phase and could be beneficial for raising the fracture toughness in the weak orientation as well. Nevertheless, even though there is one crucial or weak specimen orientation, in the other testing directions, a much higher fracture resistance was found. This means that by a rational design of components in which the anisotropy of the material is taken into account, this material type could be used for applications demanding high strength and fracture toughness.

## 5. Summary and Conclusions

In this study, the deformation and fracture behavior of a $Cu_{30}Mo_{70}$ composite subjected to high pressure torsion was investigated. To account for the microstructural alignment upon deformation, the fracture toughness was measured for different crack plane orientations. The main conclusions are as follows: 1) The HPT process not only leads to a refinement but also to a re-orientation of the individual constituents parallel to the shear plane of the disk. 2) The changes in the microstructure result in a substantial increase in hardness and strength, and even a moderate amount of ductility could be restored in the intermediate SPD regime. 3) The microstructural changes induce a pronounced anisotropy in the fracture toughness and failure behavior with weak crack paths along the microstructural alignment. This re-arrangement is however significant for the toughness enhancement through deflection and delamination toughening for the other testing directions.

## Supporting Information

Supporting Information is available from the Wiley Online Library or from the author.

## Acknowledgements

This project has received funding from the European Research Council (ERC) under the European Union's Horizon 2020 research and innovation programme (grant agreement No 340185 and No 757333). The authors thank also R. Neubauer, P. Kutleša, and M. Rockenschaub for their help with the sample preparation.

## Conflict of Interest

The authors declare no conflict of interest.

## Keywords